# Controlling Magnetism in the 2D van der Waals Antiferromagnet $CrPS_4$ via Ion Intercalation

*Alberto M. Ruiz, Diego López-Alcalá, Gonzalo Rivero-Carracedo, Andrei Shumilin, José J. Baldoví[†,*]*

A. M. Ruiz, D. López-Alcalá, G. Rivero-Carracedo, A. Shumilin, J. J. Baldoví

Instituto de Ciencia Molecular, Universitat de València, Catedrático José Beltrán 2, 46980 Paterna, Spain.

E-mail: j.jaime.baldovi@uv.es

## Abstract

Two-dimensional van der Waals (vdW) magnetic materials are versatile platforms for tailoring electronic and magnetic properties, in which the insertion of chemical species into their interlayer gaps offers a powerful route to engineer magnetism. Here, we focus on the A-type antiferromagnetic semiconductor $CrPS_4$ ($T_N$ = 38 K) and investigate its electronic and magnetic properties upon intercalation of lithium ($Li^+$) and organic tetrabutylammonium ($TBA^+$) ions using first-principles calculations. Our results show that $Li^+$ incorporation induces a semiconductor-to-metal transition in $CrPS_4$ and selectively modifies its magnetic behaviour: switching from out-of-plane to in-plane antiferromagnetism, followed by an in-plane ferromagnetic ground state at higher intercalation levels. This is accompanied by a continuous increase of the ordering temperature, reaching a fivefold enhancement for $Li_{0.5}CrPS_4$. Similarly, $TBA^+$ intercalation expands the vdW gap, decoupling $CrPS_4$ layers and stabilising in-plane ferromagnetism with a $T_C$ above 100 K. Furthermore, it also modifies magnon propagation, leading to enhanced group velocities and a more isotropic magnon transport. This work highlights intercalation as a powerful and versatile approach for controlling magnetic behaviour and spin dynamics, paving the way for the design of tunable 2D layered magnetic materials for spintronic and magnonic applications.

## 1. Introduction

Two-dimensional (2D) van der Waals (vdW) magnets have emerged as ideal platforms to explore low-dimensional magnetism and hold great promise as potential candidates for spintronics, data storage and sensing applications.[1–4] Their reduced dimensionality and high surface-to-volume ratio allow fine-tuned control of their properties under external stimuli such as strain, pressure, molecular deposition, electrostatic gating or twisting.[5–14] Among the existing 2D vdW magnetic materials, CrSBr and $CrPS_4$ stand out due to their and semiconducting character and distinct magnetic anisotropies.[15–20] $CrPS_4$, in particular, is an air-stable layered A-type antiferromagnet that has been successfully integrated into field-effect transistors, allowing gate voltage and perpendicular electric field control of its Néel temperature ($T_N$), band structure, and magnetization direction.[21–23] Moreover, long-distance magnon transport and significant modulation of thermal spin currents have been demonstrated in this system.[24–28] Owing to the weak vdW interactions between layers in these compounds, the intercalation of chemical species into the interlayer gaps offers a promising route to manipulate its magnetic behaviour.

Among the available tuning strategies, intercalation —the insertion of atoms, ions, or molecules into the vdW gaps— provides a powerful method to modulate interlayer interactions, induce charge transfer and modify the electronic structure.[29–31] This approach has enabled precise control over superconductivity, electronic transport, and magnetic properties by inducing interlayer expansion, charge transfer between host and guest species, orbital hybridization, or phonon scattering.[32–34] For instance, Cu intercalation in $2H-NbS_2$ enhances electrical conductivity while suppressing superconductivity,[35] and K intercalation in $2H-MoS_2$ triggers a structural phase transition to the 1T and 1T' phases, leading to superconducting behavior.[36] In vdW magnetic materials, $Li^+$ intercalation in $CrI_3$ boosts its Curie temperature ($T_C$),[37] whereas in CrSBr it simultaneously increases electrical conductivity, magnetic ordering temperature, and drives an antiferromagnetic-to-ferromagnetic phase transition.[38–40] Beyond alkali metals, organic cations introduce additional tunability due to their structural complexity, enabling both control of doping level and selective modulation of interlayer spacing.[41,42] Specifically, intercalation of organic cations in NiPS$_3$ decouples the magnetic layers, induces ferrimagnetic order, and shifts its T$_C$.[43] Furthermore, the incorporation of tetrabutylammonium (TBA$^+$) in Cr$_2$Ge$_2$Te$_6$, Fe$_3$GeTe$_2$ and CrSBr enhances their ordering temperatures, with Cr$_2$Ge$_2$Te$_6$ also exhibiting a reorientation of its magnetization axis.[39,44,45]

Despite the demonstrated potential of ion intercalation to tune the properties of 2D vdW magnets, its impact on CrPS$_4$ remains largely unexplored. In this work, we investigate the effects of Li$^+$ and organic TBA$^+$ intercalation in CrPS$_4$ using first-principles calculations. We show that these guest species modulate the electronic structure and selectively tune intralayer and interlayer exchange interactions, leading to enhanced magnetic ordering temperature, reoriented magnetization easy axis direction and modified magnon transport. Our results establish intercalation as a versatile design strategy for fine-tuning the electronic and magnetic properties in 2D vdW magnets.

2. Results and Discussion

CrPS$_4$ crystallizes in a monoclinic structure with space group $C_2$ (no. 5), where Cr and P atoms are coordinated in a distorted octahedral and tetrahedral environments, respectively. The octahedral crystal field around the Cr atoms splits the d orbitals into lower-energy t$_{2g}$ (d$_{xy}$, d$_{xz}$, and d$_{yz}$) and higher-energy e$_g$ (d$_{z2}$ and d$_{x2-y2}$) states.[20,46] Magnetically, CrPS$_4$ exhibits an A-type antiferromagnetic (AF) ground state, composed of ferromagnetic (FM) layers coupled antiferromagnetically along the c-axis (Figure 1a), with a Néel temperature of T$_N$ = 38 K.[18] To gain insight into the electronic and magnetic properties of bulk CrPS$_4$, we perform first-principles calculations within the DFT+U framework, including a Hubbard correction to account for on-site Coulomb interactions. The optimized lattice parameters ($a$ = 10.89 Å, $b$ =

7.29 Å, $c$ = 6.13 Å, and $\beta$ = 92°) are in good agreement with prior studies.[47,48] The computed magnetic moment for each Cr atom is 2.92 $\mu_B$, consistent with the expected S = 3/2 for $Cr^{3+}$ ions and closely matching the experimentally value of 2.8 $\mu_B$.[18] Band structure calculations confirm the semiconducting behaviour of $CrPS_4$ (Figure 1c), yielding an estimated band gap of 0.87 eV. This value agrees well with scanning tunneling spectroscopy (STS) measurements and previous theoretical results.[21,49] The gap arises from a spin-split band structure, with the valence band maximum and conduction band minimum showing opposite spin character (Figure S1).[46,48–50] Additionally, a spin-conserving transition between bands of the same spin character leads to a larger band gap, in agreement with photoluminescence (PL) measurements, and is attributed to spin-allowed d–d transitions involving Cr $t_{2g}$ and $e_g$ orbitals.[51–53] Orbital-resolved band structure analysis (Figure 1c) indicates that the Cr d orbitals dominate the electronic states near the Fermi level, accompanied by substantial contributions from p orbitals of S atoms.

The magnetic behaviour of $CrPS_4$ is governed by five dominant intralayer magnetic exchange interactions, $J_1$–$J_5$ (Figure 1b). These are extracted by constructing a tight-binding Hamiltonian based on maximally localized Wannier functions (see Supporting Information), yielding $J_1$ = 2.85 meV, $J_2$ = 2.59 meV, $J_3$ = 0.05 meV, $J_4$ = 1.11 meV and $J_5$ = -0.83 meV, where positive values denote FM coupling. The larger magnitude of $J_1$ relative to $J_2$ arises from the shorter Cr–Cr distance (3.59 vs. 3.70 Å) and the Cr–S–Cr bond angle close to 90° (94.98° vs. 98.07°), favouring FM superexchange. In contrast, interlayer couplings are AF but considerably weaker than the dominant intralayer FM interactions (Figure S2; Tables S1 and S2, Supporting Information). To quantify the effective interlayer exchange ($J_{int}$), we calculate the total energy difference between AF and FM spin configurations, defined as $J_{int} = E_{AF}-E_{FM}$. We estimate $J_{int}$ = –1.06 meV/Cr, which correctly reproduces the A-type AF ground state of bulk $CrPS_4$.

This result agrees well with the reported experimental value of –0.58 meV/Cr for bilayer CrPS$_4$, where the twofold enhancement in the bulk relative to the bilayer is consistent with their different spin-flip fields of 7–8 T and 3.5 T, respectively.[18,21,22]

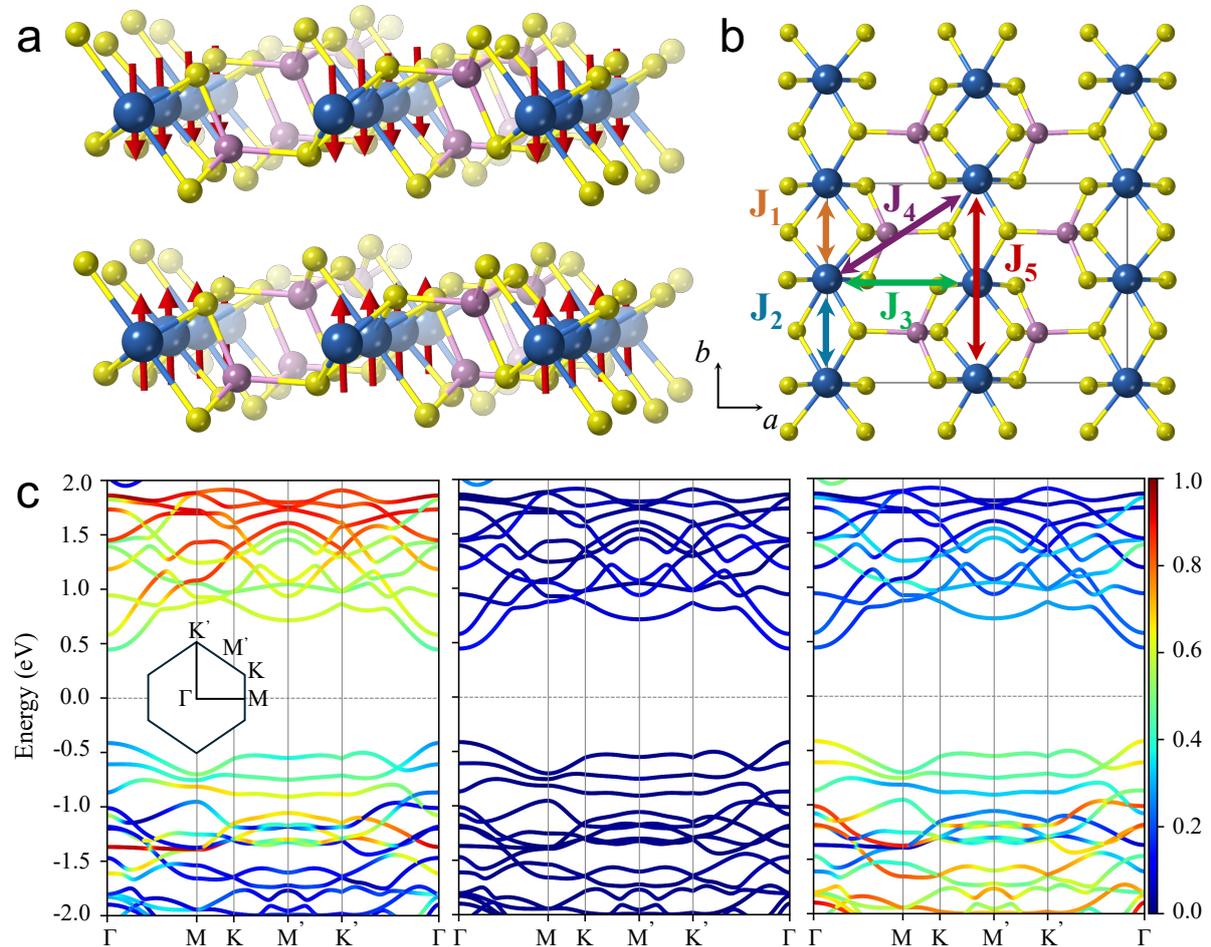

**Figure 1.** Structural, electronic, and magnetic properties of bulk and single-layer CrPS$_4$. a) Lateral view of bulk CrPS$_4$ exhibiting AF ground state. Colour code: Cr (blue), P (pink) and S (yellow). b) Top view of single-layer CrPS$_4$ indicating the intralayer magnetic exchange interactions $J_1$-$J_5$. c) Orbital-resolved band structure of bulk CrPS$_4$ showing the partial contribution of Cr (d orbitals), P (p orbitals) and S (p orbitals), from left to right, respectively. The colour bar indicates the normalized relative contribution to the electronic bands of each orbital.

To evaluate the magnetic anisotropy of CrPS$_4$, we consider two primary contributions: (i) the spin–orbit coupling term (SOC-MAE), and (ii) the shape anisotropy (shape-MAE), which arises from long-range dipole–dipole interactions. While shape-MAE is typically weaker than SOC-MAE in most 2D magnets, it becomes non-negligible in systems such as CrPS$_4$ or

CrSBr,[54] where the orbital moment of $Cr^{3+}$ ions is quenched due to half-filling of the low-energy $t_{2g}$ orbitals. Considering contributions of both SOC-MAE and shape-MAE, we obtain $MAE_{bc} = E_b – E_c = 24.5$ μeV/Cr and $MAE_{ac} = E_a – E_c = 39.8$ μeV/Cr (see Table S3, Supporting Information). Our results reveal a preferential out-of-plane spin orientations, where $b$ and $a$ axes are the intermediate and hard magnetization directions, respectively.[18] Atomistic spin simulations based on the extracted exchange couplings and magnetic anisotropy yield a $T_N = 39$ K, in excellent agreement with the experimental result of 38 K, thus validating the theoretical framework.[18]

Then, we investigate the intercalation of $Li^+$ ions into the vdW gap of $CrPS_4$ (Figure 2a) by considering compositions $Li_{0.125}CrPS_4$, $Li_{0.25}CrPS_4$, $Li_{0.375}CrPS_4$, and $Li_{0.5}CrPS_4$. Higher concentrations are not studied, as it represents the upper limit in related layered materials such as $FePS_3$ and $NiPS_3$, beyond which structural degradation and eventual breakdown of the samples have been observed.[55,56] The stability of $Li^+$ intercalation is evaluated at multiple sites within the vdW gap (Figures S7 and S8, Supporting Information), and in all cases, the intercalation is energetically favourable (Table S5, Supporting Information) based on the results of adsorption energy ($E_{ads}$). Among them, $Li^+$ preferentially occupies positions above the Cr–Cr hollow along the $J_1$ direction (Figure S7, Supporting Information), showing a value of $E_{ads} = -7.02$ eV. Note that the Li-S distance is ~2.5 Å, significantly shorter than the typical 3–4 Å vdW spacing, pointing to chemical bonding rather than weak vdW interactions. This is further supported by manually increasing the Li-S separation along the $c$ direction, resulting in a less stable configuration (Figure S9, Supporting Information).

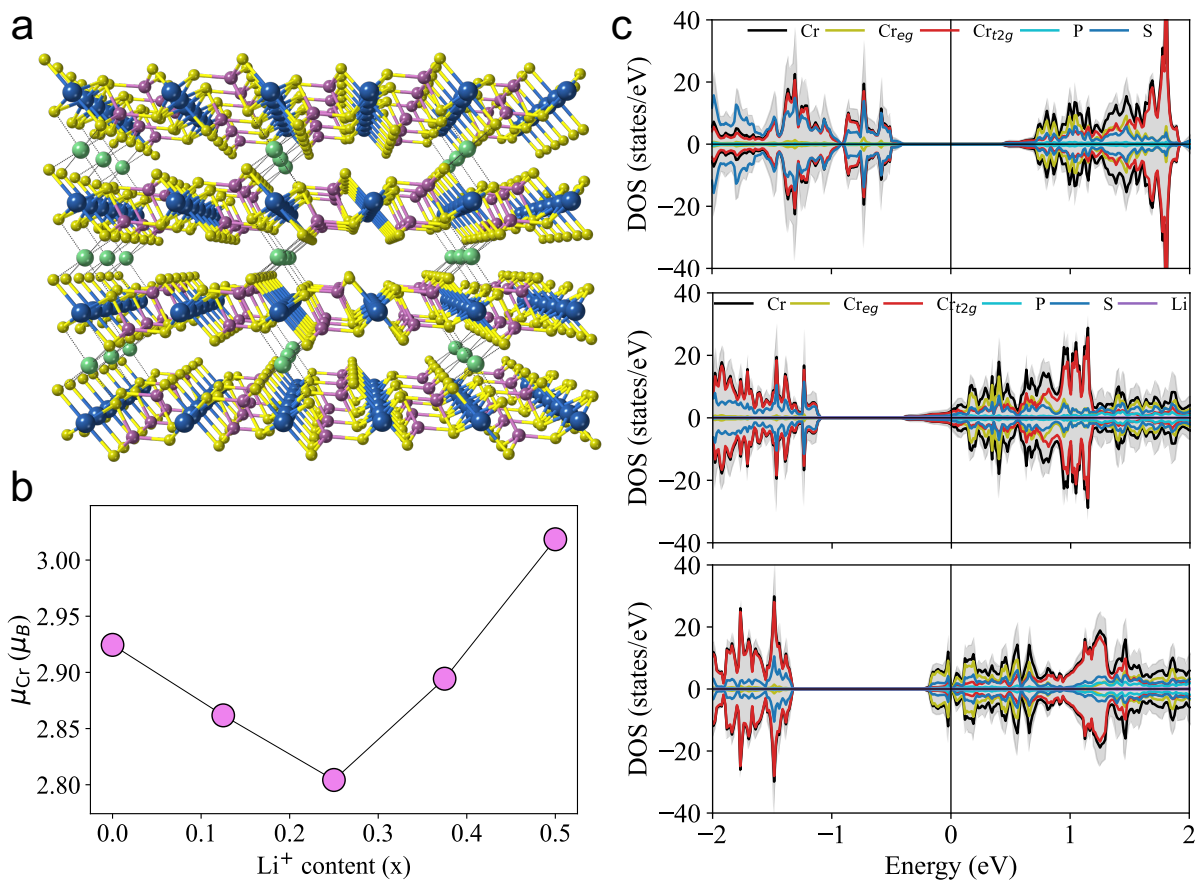

**Figure 2.** Structural, magnetic, and electronic properties of Li-intercalated $CrPS_4$. a) Lateral view of $Li_{0.25}CrPS_4$, where green balls depict $Li^+$ ions. b) Evolution of magnetic moments of Cr atoms upon increasing the $Li^+$ content. c) Orbital resolved DOS for $CrPS_4$, $Li_{0.25}CrPS_4$ and $Li_{0.5}CrPS_4$ (from top to bottom, respectively), where the grey filling represents the total DOS.

Upon $Li^+$ intercalation, both in-plane lattice parameters as well as the interlayer spacing undergo a progressive expansion (Figure S10, Supporting Information), reaching a maximum increase of 1.6%, 1.7% and 1.6% for *a*, *b* and *c* lattice parameters, respectively, for $Li_{0.5}CrPS_4$. Conversely, the average Cr magnetic moment exhibits a non-monotonic evolution, decreasing up to $Li_{0.25}CrPS_4$, and then increasing for $Li_{0.375}CrPS_4$ and $Li_{0.5}CrPS_4$ (Figure 2b). This non-monotonic trend indicates that for $Li^+$ content ≤ 0.25, the electrons transferred to $CrPS_4$ do not reduce $Cr^{3+}$ ($d^3$) to $Cr^{2+}$ ($d^4$),[57] which would otherwise enhance the Cr magnetic moments via partial occupation of the empty $e_g$ orbitals. This is confirmed by calculating the orbital-resolved density of states (DOS), which shows that in pristine $CrPS_4$ the Fermi level resides within the gap, typical of a semiconducting behaviour, with unoccupied Cr $t_{2g}$ and S p orbitals located just above the gap (Figure 2c). In $Li_{0.25}CrPS_4$, the Fermi level shifts into the conduction band, inducing a metallic state. The additional electrons occupy Cr $t_{2g}$ orbitals with opposite spin polarization relative to those at the valence band maximum (Figure S11, Supporting Information), leading to a net reduction of the Cr magnetic moment. For $Li_{0.5}CrPS_4$, spin up Cr $e_g$ states are occupied, resulting in the observed increase in magnetic moments at higher $Li^+$ concentrations.

The evolution of interlayer exchange coupling ($J_{int}$) reveals a robust AF state up to $Li_{0.29}CrPS_4$ (Figure 3a) while higher $Li^+$ contents ($Li_{0.29}CrPS_4$–$Li_{0.5}CrPS_4$) stabilize ferromagnetism through the filling of Cr $e_g$ orbitals. Additionally, the out-of-plane magnetic anisotropy softens progressively, driving a reorientation to in-plane magnetisation with spins pointing along the *a* axis for $Li^+$ contents $\geq 0.17$. Note that for higher $Li^+$ concentrations ($Li_{0.4}CrPS_4$–$Li_{0.5}CrPS_4$), the *b* axis becomes the intermediate magnetization direction, while the *c* axis is the hard magnetic state (Figures S12 and S13, Supporting Information). This defines a phase diagram with three different regions: (i) an AF out-of-plane state resembling pristine $CrPS_4$ at low $Li^+$ content (green), (ii) an AF in-plane state for $Li_{0.17}CrPS_4$–$Li_{0.29}CrPS_4$ (blue), and (iii) an ultimate in-plane FM configuration for higher $Li^+$ concentrations emerging upon occupation of Cr $e_g$ orbitals (red). From the evolution of the intralayer $J_1$-$J_5$ exchange couplings one can observe that all are enhanced upon intercalation (Figure 3b). This can be attributed to the additional electrons introduced by $Li^+$ ions, leading to a significant strengthening of magnetic couplings.[44] As a result, the magnetic ordering temperature increases progressively from $T_N$ = 39K in pristine $CrPS_4$ to a maximum of $T_C$ = 228K for $Li_{0.5}CrPS_4$ (Figure 3c).

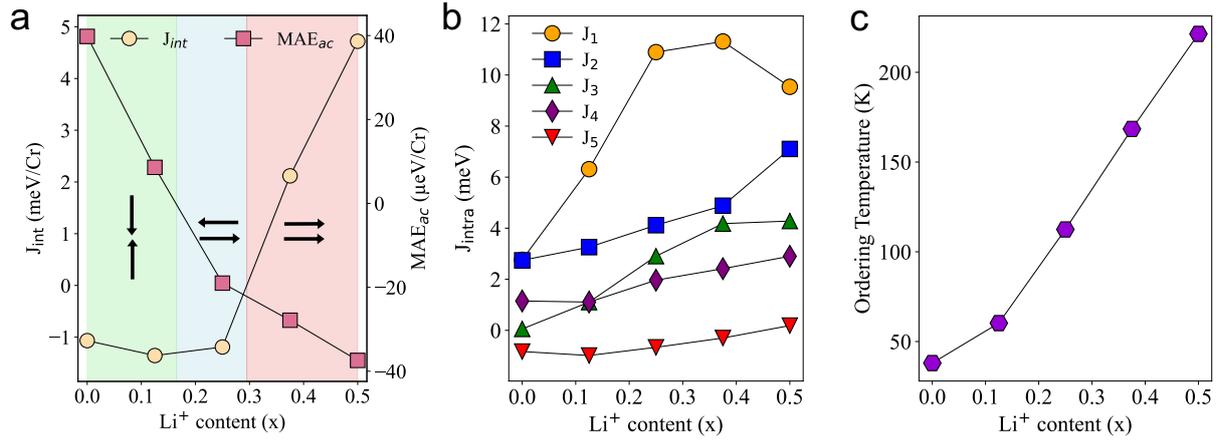

**Figure 3.** Evolution of the exchange interactions, magnetic anisotropy and ordering temperature for $Li_xCrPS_4$. a) Evolution of interlayer exchange interaction ($J_{int}$) and MAE, where the black arrows indicate the magnetic ground state, b) intralayer couplings $J_1$-$J_5$ and c) magnetic ordering temperature for $Li_xCrPS_4$.

To gain deeper insights into the mechanisms underlying the observed magnetic behaviour, we obtain the evolution of the magnetic properties of $CrPS_4$ as a function of electron doping. This approach allow us to decouple the effects of (i) Fermi level shifts driven by the additional electrons from (ii) the structural distortions induced by $Li^+$ intercalation. Figure S14 shows that electron doping initially reduces the Cr magnetic moments, keeping $J_{int}$ nearly constant. Upon occupation of the $e_g$ orbitals the magnetic moments increase, rapidly stabilizing interlayer ferromagnetism at 0.42 $e^-$/f.u. (Figure S15, Supporting Information). Note that the stabilization of the interlayer ferromagnetism occurs independently of the chosen Hubbard U, resulting in an earlier FM ground state for larger U values (Figure S15, Supporting Information). Since each $Li^+$ ion ideally donates one electron to $CrPS_4$ (e.g., $Li_{0.5}CrPS_4$ transfers 0.5 $e^-$/f.u.), higher electron densities are required to induce interlayer ferromagnetism via electrostatic doping than via $Li^+$ intercalation (0.42 $e^-$/f.u. vs 0.29 $e^-$/f.u., respectively). This difference arises from the expansion of the interlayer spacing upon $Li^+$ incorporation, which weakens the AF interlayer coupling and reduces the number of electrons needed to stabilize the FM state. The added carriers also gradually reduce the out-of-plane magnetic anisotropy, ultimately favouring an in-plane magnetic configuration with enhanced ordering temperature (Figures S16-S18, Supporting Information).

Then, we investigate the effect of intercalating a guest molecular cation, specifically the TBA$^+$ ion. Two intercalation levels are considered, namely (TBA)$_{0.125}$CrPS$_4$ and (TBA)$_{0.25}$CrPS$_4$, corresponding to concentrations that have been experimentally achieved in (TBA)$_{0.25}$NiPS$_3$.[43] Note that the intercalation induces a pronounced expansion of vdW gap, with the interlayer spacing increasing from 2.5 Å in pristine CrPS$_4$ to 14.2 Å in the intercalated compounds (Figure 4a). Concurrently, a slight compression of the in-plane lattice parameters is observed, resulting in values of $a$ = 10.82 Å and $b$ = 7.24 Å for (TBA)$_{0.125}$CrPS$_4$, and $a$ = 10.75 Å and $b$ = 7.20 Å for (TBA)$_{0.25}$CrPS$_4$. We perform charge density difference (CDD) calculations to investigate the charge redistribution across the heterostructure (Figure 4a). Our results indicate electron depletion primarily at the terminal hydrogen atoms of TBA$^+$ cations, whereas electron accumulation is localized on adjacent S atoms of CrPS$_4$, consistent with their higher electronegativity. Additionally, each TBA$^+$ molecule donates 0.92 e$^-$ to the substrate, as revealed by Bader charge transfer analysis, inducing a metallic behaviour through partial occupation of previously empty conduction bands (Figure S19, Supporting Information), which in turn modulate the magnetic moments of Cr atoms. Specifically, these are 2.85 μ$_B$/Cr and 2.79 μ$_B$/Cr for TBA$^+$ contents of 0.125 and 0.25, respectively, compared to 2.92 μ$_B$/Cr in pristine CrPS$_4$. This variation mirrors the prior observations in Li-intercalated CrPS$_4$ and is attributed to the occupation of empty t$_{2g}$ states due to charge transfer (Figure S19, Supporting Information).

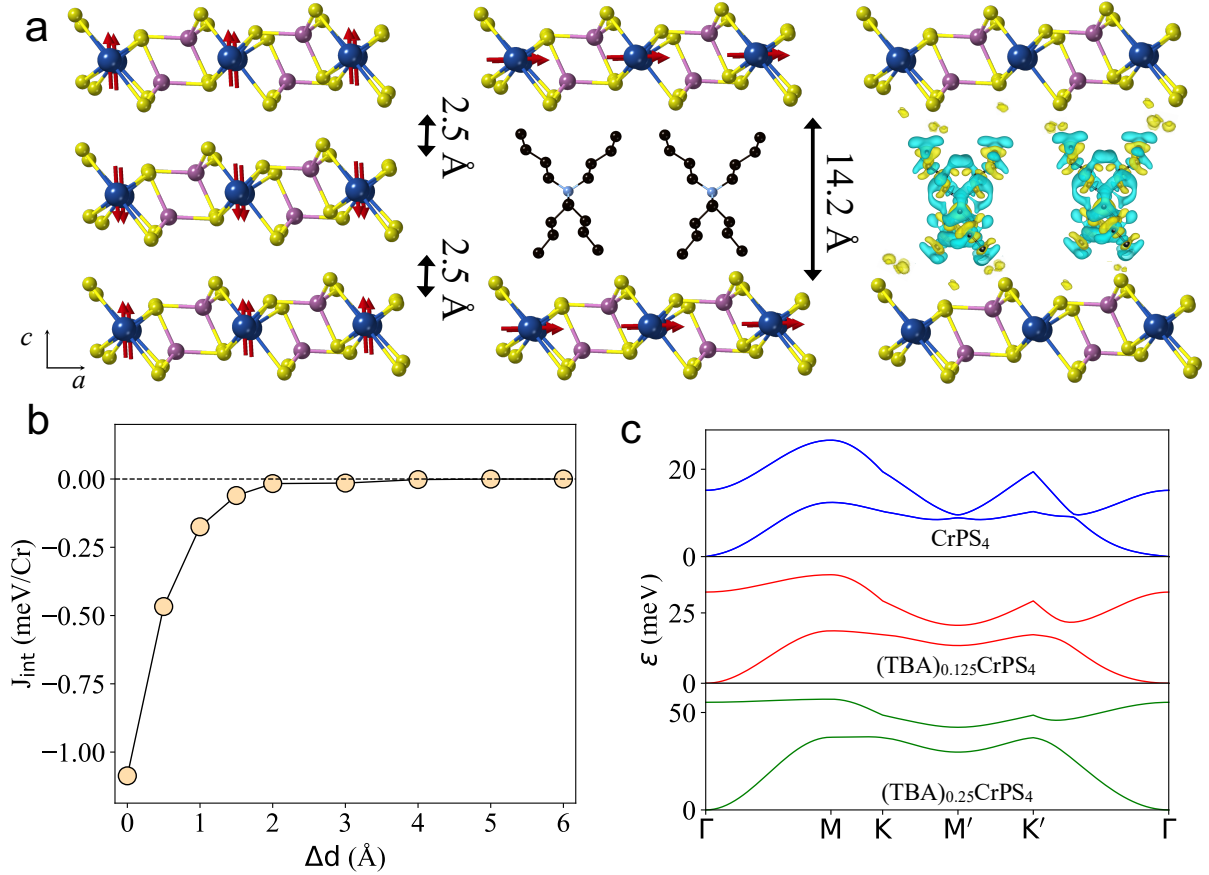

**Figure 4.** Structural and magnetic properties as well as magnon dispersion of $(TBA)_xCrPS_4$. a) Side views of pristine $CrPS_4$ and $(TBA)_{0.25}CrPS_4$, showing the interlayer spacing and magnetization orientation in each case together with charge density difference (CDD) plots for the latter, where yellow (blue) denotes electron accumulation (depletion). b) Evolution of $J_{int}$ as a function of the change in interlayer distance ($\Delta d$) for pristine $CrPS_4$. c) Magnon dispersion for pristine $CrPS_4$, $(TBA)_{0.125}CrPS_4$ and $(TBA)_{0.25}CrPS_4$, from top to bottom, respectively.

$TBA^+$ incorporation effectively decouples the $CrPS_4$ layers along the $c$ axis, resulting in magnetically isolated layers each one exhibiting a FM ground state. To illustrate this effect, we compute the variation of $J_{int}$ for pristine $CrPS_4$ as a function of the change in interlayer distance ($\Delta d$). As shown in Figure 4b, the AF coupling rapidly decreases with increasing distance and saturates at $\Delta d \approx 3$ Å. Since the interlayer separation in $(TBA)_xCrPS_4$ exceeds this value ($\Delta d =$ 11.7 Å), $J_{int}$ is effectively suppressed. Interestingly, we observe that in the intercalated compounds the magnetization is switched to the in-plane $a$ direction, stabilised by the transferred electrons from $TBA^+$ to the substrate. Particularly, $MAE_{ac}$ changes from

39.8 µeV/Cr (CrPS$_4$) to -25.4 µeV/Cr and -48.6 µeV/Cr for TBA$^+$ contents of 0.125 and 0.25, respectively (Table S6, Supporting Information). Upon intercalation, the FM intralayer exchange interactions are enhanced (Table S7, Supporting Information), resulting in T$_C$ of 51 K for (TBA)$_{0.125}$CrPS$_4$ and 104 K for (TBA)$_{0.25}$CrPS$_4$.

Within 2D magnets, CrPS$_4$ has been at the forefront of research on magnon excitations,[24,58,59] including experimental observations of magnon propagation over large distances of ~1 µm.[24,59] The magnon spectrum of CrPS$_4$ exhibits four doubly degenerate magnon bands (Figure 4c; Figure S20, Supporting Information). Notably, at the M' point and along Γ–K' direction, the acoustic and optical branches are nearly degenerate due to the near equality of the exchange interactions J$_1$ (2.85 meV) and J$_2$ (2.59 meV). Furthermore, the energy of acoustic magnons increase more rapidly along the y direction (Γ– K') than along x (Γ–M), resulting in maximum group velocities of $v_y$ = 8.4 × 10$^3$ m.s$^{-1}$ and $v_x$ = 4.9 × 10$^3$ m.s$^{-1}$, respectively, in agreement with previous findings.[28] This anisotropy arises from the larger values of J$_1$ and J$_2$, which are aligned along y, compared to J$_3$ (0.05 meV) along x, while J$_4$ contribute to both as it points along the diagonal direction. Intercalation with TBA$^+$ molecules lifts the degeneracy at M' and Γ–K' due to the marked differences between J$_1$ and J$_2$ (Table S7, Supporting Information) and results in enhanced magnon velocities. For (TBA)$_{0.125}$CrPS$_4$, $v_y$ and $v_x$ reach 9.9 × 10$^3$ m.s$^{-1}$ and 7.7 × 10$^3$ m.s$^{-1}$, respectively, further increasing to 13.9 × 10$^3$ m.s$^{-1}$ and 15.4 × 10$^3$ m.s$^{-1}$ for (TBA)$_{0.25}$CrPS$_4$ (Figure S21). This indicates that TBA$^+$ cations not only increase magnon velocities but also make their propagation more isotropic, primarily due to the enhanced magnitude of J$_3$ (Table S7, Supporting Information).

Our results highlight molecular intercalation in 2D vdW magnetic materials as a powerful and versatile tool to chemically control magnonic devices. This combines the accessibility of experimental techniques to measure magnons in bulk systems (e.g. magnon dispersion via inelastic neutron scattering)[60,61] with the interfacial tunability of 2D layered materials,[62] enabling precise layer-by-layer modulation of the magnetic properties of the system.

## 3. Conclusion

In summary, we have investigated the tunability of the electronic and magnetic properties of $CrPS_4$ via ion intercalation using first-principles calculations. We demonstrate that $Li^+$ intercalation induces a semiconductor-to-metal transition while selectively modifying magnetic exchange interactions, giving rise to a sequence of magnetic phases: an out-of-plane antiferromagnetic state at low $Li^+$ content, an in-plane antiferromagnetic regime at intermediate concentrations, and a ferromagnetic in-plane ground state at higher intercalation levels. This is rationalized from a microscopic analysis due to the different occupations of $CrPS_4$ orbitals upon electron doping. Additionally, $Li^+$ intercalation leads to a significant enhancement of the magnetic ordering temperature, reaching $T_C = 228$ K for $Li_{0.5}CrPS_4$, driven by the strengthening of intralayer exchange interactions. On the other hand, the intercalation of organic $TBA^+$ cations expand the vdW gap, effectively decoupling $CrPS_4$ layers. Notably, the transferred electrons from $TBA^+$ modify the electronic structure, inducing metallicity. Furthermore, $TBA^+$ intercalation stabilizes in-plane ferromagnetism, resulting in an enhanced $T_C$ above 100K, increased magnon group velocities and a more isotropic magnon transport. Our work provides a microscopic understanding of how chemical species inserted into the interlayer gaps influence electronic and magnetic properties, offering a pathway to design highly tunable 2D magnetic materials for spintronics and magnonic applications.

## 4. Methods

Spin polarized density-functional theory (DFT) calculations for $CrPS_4$ were performed using the VASP package.[63] The exchange-correlation energy was treated within the generalized gradient approximation (GGA). To accurately describe the electronic and magnetic properties of CrPS$_4$, a Hubbard-corrected DFT+U approach was employed with a Hubbard U value of 0.25 eV. This value of U which was chosen to best reproduce the experimental electronic band gap and T$_N$ (see Supporting Information) and correctly captures the electronic properties obtained using the HSE06 hybrid functional. We employed the DFT-D2 to describe the vdW interactions between adjacent layers of CrPS$_4$. Maximally localized Wannier functions were constructed using Wannier90,[64] with the d orbitals of Cr and the p and s orbitals of P and S as the basis, to generate a tight-binding Hamiltonian. Magnetic exchange couplings were subsequently calculated using TB2J code, employing supercells of dimensions 10 × 10 × 5.[65] T$_N$ and T$_C$ were determined through atomistic simulations using the VAMPIRE code.[66] Simulations were performed on supercells of dimensions 25 nm × 25 nm × 25 nm, with both equilibration and averaging phases performed over 10,000 steps employing the llg-heun integrator.

## Acknowledgements

The authors acknowledge the financial support from the European Union (ERC-2021-StG-101042680 2D-SMARTiES) and the Generalitat Valenciana (grant CIDEXG/2023/1). A.M.R. thanks the Spanish MIU (Grant No FPU21/04195) and G.R.C. thanks the University of Valencia (grant Atracció de Talent INV23-01-13). The calculations were performed on the HAWK cluster of the 2D Smart Materials Lab hosted by Servei d'Informàtica of the Universitat de València.